\documentclass[prd,aps,
superscriptaddress,preprintnumbers,nofootinbib,showpacs]{revtex4}

\usepackage{graphicx}
\usepackage{exscale}
\usepackage[intlimits]{amsmath}
\usepackage{amsfonts}
\usepackage{amssymb,amscd}
\usepackage{epsfig}
\usepackage[english]{babel}
\usepackage{amsfonts,amsmath,amssymb}
\usepackage{float,braket}
\usepackage{pslatex}
\usepackage{graphicx,color}
\usepackage{subfigure}
\usepackage{graphicx,wrapfig}
\usepackage[utf8]{inputenc}

\begin{document}

\title{Local QCD action at Finite Temperature}

\author{Patrick~Cooper}
\email[]{pjc370@nyu.edu}
\affiliation{New York University, New York, NY 10003}

\author{Daniel~Zwanziger}
\email[]{dz2@nyu.edu}
\affiliation{New York University, New York, NY 10003}

\date{Dec. 29, 2015}

\begin{abstract}
        In this article, we obtain a local, BRST-invariant action for QCD at
        finite temperature, altered from Fadeev-Popov theory due to the presence of
        Gribov copies. We carefully derive the horizon condition at finite
        temperature. Only the zero Matsubara mode is affected, and this result
        is consistent with the suitably modified Maggiore-Schaden shift, which
        takes into account temporal periodicity. The large-N limit and other
        calculational schemes for the magnetic mass and gluon condensates and
        their relation to the Gribov mass are also discussed. 
\end{abstract}

\maketitle

\section{Introduction} %

Quantum Chromodynamics (QCD) at high temperatures is one corner of the standard
model that that is still wrought with mystery, and is an active area of
research. Questions involving the very early universe, as well as the interior
of neutron stars are both limited by our lack of understanding of the
hypothesized Quark Gluon Plasma (QGP), presumed to exist at high temperatures.
Intuitively, due to asymptotic freedom one would expect that at high
temperatures, typical momentum transfers are large so quarks should behave like
free particles. From the lattice data \cite{Karsch:2003jg}, the equation of state
does approach a Stefan-Boltzmann limit, supporting this hypothesis. There's
another theory, however, that also shows similar ideal-gas behavior at
high temperatures: $\mathcal{N}=4$ super Yang-Mills in the Large $N$ limit.
This theory, while integrable, is very strongly coupled and doesn't behave
simply like free particles \cite{Arnold:2007pg}.

Another sign that this analysis of finite-temperature QCD may be more subtle
than our naive intuition is that in the high-temperature limit, the time
direction becomes vanishingly small and thus decouples and the action 
dimensionally reduces \cite{Arnold:2007pg}
\begin{equation}
\label{vanishinglysmall}
        S \; \; = \; \; \frac{1}{\hbar} \int_0^{\beta \hbar} d\tau \int d^dx \,
        \mathcal{L}_{QCD} \; \; \rightarrow \; \; \beta \int d^dx \,
        \mathcal{L}_{QCD} ,
\end{equation}
where $D = d+1$ is the total spacetime dimension with $d$ the number of spatial
dimensions. Thus the high-temperature limit of a quantum field theory becomes
classical\footnote{Intuitively this is due to the fact that at high
        temperatures, occupation numbers are high which corresponds to the
classical limit} in a sense (no more $\hbar$) in one less dimension. We know
however, due originally to Feynman \cite{Feynman:1981ss}, that 3-D QCD is a
confining theory. Thus if we took asymptotic freedom at face value, a 4-D
theory of free particles would parametrize a confining 3 dimensional field
theory -- an odd proposition.

On a much more concrete level, QCD is plagued by infrared divergences in ways
that its abelian cousin, Quantum Electrodynamics (QED) is not. The most
important object to calculate in statistical mechanics is the partition
function. From this object, all thermodynamic quantities of interest can be
obtained by merely calculating the gradients of this function along
various directions in parameter space. In QED, one can calculate the partition
function perturbatively to arbitrarily high powers in the coupling, $\alpha$.
In QCD, no such construction has been found. While non-perturbative information
is similarly obtained by summing an infinite number of so-called ring diagrams
\cite{Kapusta:2006pm}, and a Debye mass is dynamically generated for $A_0$ not
unlike QED, one quickly runs into trouble at higher orders due to the non-linearity of
gluon-gluon scattering. By adding more and more gluon loops to a perturbative calculation of the free energy, one obtains diagrams of the same order, $g^6$, in the Yang Mills
coupling constant \cite{Gross:1980br}. This also similarly occurs at order
$g^4$ when calculating higher order contributions to the gluon self energy.
This phenomenon was originally pointed out by Linde in 1979 \cite{Linde:1978px}.

Since many issues of finite-temperature QCD seem to be related to its
infrared behavior (as with most mysteries about QCD) it seems logical that a clue
for how to amelioriate this problem may be related to an inconsistency with our
formulation of gauge theory that is very prominent in the infrared, namely the
Gribov Ambiguity.

In 1978, Gribov published his famous paper \cite{Gribov:1977wm} where he showed
that the Fadeev Popov (FP) method of gauge fixing was incomplete. He did this
by inferring that zero modes of the FP operator imply that each transverse gauge
connection has copies belonging to the same gauge orbit.
Singer shortly thereafter \cite{singer1978} showed that finding a gauge fixing
condition that would lead to a globally defined, continuous gauge condition
which behaved well at infinity was generically impossible for non-abelian gauge
theories due to the highly non-trivial topology of the space of gauge orbits.

By defining a region that drastically reduced this redundancy, Gribov worked
out the dynamical consequences that the elimination of these copies could have. What he found led to clues piercing into the heart of confinement and kicked off a now mature,
four-decade old research program. Of interest to us, is that his alterations to
the configuration space of gauge connections affects the infrared sector of the
theory dramatically \cite{Gribov:1977wm}: the gluon propagator now vanishes at
$k=0$ due to the presence of an infrared regulator known as the ``Gribov mass"
and the pole is removed from the real line. The ghost propagator on the other
hand obtains an enhanced singularity which, as is shown in an accompanying
paper \cite{Cooper:2015}, is related to the divergence of the color-Coulomb
potential. This same enhancement was shown by Reinhardt
\cite{PhysRevLett.101.061602} to imply that the QCD vacuum is a dual color
superconductor, thus expelling electric flux the way that traditional
superconductors expel magnetic flux.

In the following sections, we'll see how contributions to the
finite-temperature partition function for Yang Mills theory restricted to the
Gribov region should be constructed. We'll find that it's not as simple as just
imposing temporal periodicity. We'll also see that this subtlety reflects ones
inability to perform the Maggiore-Schaden shift with a periodic time
coordinate. In the last sections we'll review attempts at calculating
contributions to thermodynamic quantities in the presence of the Gribov mass.
We'll discuss recent attempts to understand the magnetic mass and what role the
Gribov mass may play in this story. We'll find that the local GZ action at
finite temperature doesn't affect classical solutions to the gauge field. 

Finally in an appendix we show that at high temperature, one can carry out
the calculation of the horizon condition perturbatively in the zero-temperature,
dimensionally-reduced euclidean theory and this will yield results consistent with
first taking the sum over Matsubara frequencies and then taking the high-temperature limit. This shows that the gap equation at high temperature is
the same as at zero temperature in one less dimension, consistent with \eqref{vanishinglysmall}. 

\section{Reviewing the Gribov Horizon at Zero Temperature}%

Since Gribov's '78 paper, a multitude of results and observations have been
achieved from attempting to cleanse the configuration space of gauge copies that
survive after covariant gauge fixing.\footnote{For a comprehensive list of
        references up to 2012 on this topic, see the references at the end of
        the review \cite{Vandersickel:2012tz}.} The Gribov
region, $\Omega$, is defined to be the set on which the FP operator,
$\mathcal{M}(A)^{ab}(x,y) \equiv  (- \partial \cdot D)^{ab}(x) \delta(x-y)$, is
positive definite:
\begin{equation}
        \Omega = \{ A_{\mu}^a : (\theta^a,\mathcal{M}(A)^{ab} \theta^b) > 0 \;
        , \; \forall \theta \}
\end{equation}
where $\theta$ is an arbitrary $L^2$ function, and the inner product is the
usual inner product on $L^2$ space. The FP ghost kinetic term is precisely of
this form so Gribov began his analysis by doing a semi-classical
calculation to investigate where the 1PI ghost propagator crossed from positive
to negative, away from the usual pole at $k = 0$. This is the so-called ``no-pole"
condition. Since then, others \cite{Gomez:2009tj} have refined this technique to
obtain the horizon function from the no-pole condition.  This is the functional of $A_{\mu}^a$,\footnote{Shown here in Landau gauge}
\begin{equation}
        H(A) \equiv \int d^Dx d^Dy \; D_{\mu}^{ab}(x) D_{\mu}^{ac}(y)
        [(\mathcal{M}(A))^{-1}]^{bc}(x,y) \; , \nonumber
\end{equation}
which
characterizes the Gribov region $\Omega$ by the condition $H(A) \leq (N^2-1)dV$
The surface in configuration space at which $H(A) - (N^2-1)dV$ goes from positive to
negative is known as the ``Gribov horizon", $\partial \Omega$. It was shown in
\cite{Zwanziger:1982na} that this region is bounded, convex, contains the
perturbative region near $A_{\mu}^a = 0$ and is intersected by every gauge orbit.
Then, borrowing an analogy from statistical mechanics when enforcing the
constant energy constraint in the canonical ensemble\footnote{This argument
        wouldn't apply when the space of connections is a one dimensional
        integral as in Coulomb gauge in 1+1 dimensions. In this gauge in 1+1
        dimensions the Horizon condition factorizes and only contributes a
        constant to the action and the restriction to the Gribov region becomes
vacuous}, one can add this non-local constraint to the action. Akin to
requiring that the enemble average yields the desired energy, the
non-perturbative constraint to restrict configuration space to the Gribov
region is given by \begin{equation}
        \label{condition}
        \langle H(A) \rangle = (N^2-1) D V
\end{equation}
From here, the non-local term was localized by adding a BRST quartet of new
ghost fields, $\phi, \; \bar{\phi}, \; \omega, \; \bar{\omega}$, leading to the
Gribov-Zwanziger (GZ) action. It was shown that this action was
renormalizable \cite{Zwanziger:1989mf, Maggiore:1993wq, Dudal:2010fq} and
contained many new symmetries, including a spontaneously broken BRST, and it is conjectured that this
allows one to define the space of physical states in Hilbert space
\cite{Schaden:2014bea,Schaden:2015uua}. For the sake of completeness, the form
of the final GZ action (here expressed in Landau gauge) is given by 
\begin{eqnarray}
\label{gz}
\mathcal L_{\rm GZ} &= \frac{1}{4} F^2_{\mu \nu} + i \partial_{\mu} b \cdot
        A_{\mu} - i \partial_{\mu} \bar{c}^a ( D_{\mu} c)^a + \partial_{\mu}
        \bar{\phi}^a_{\nu b} (D_{\mu} \phi_{\nu b})^a - \partial_{\mu}
        \bar{\omega}^a_{\nu b} \cdot (D_{\mu} \omega_{\nu b} + D_{\mu} c \times
        \phi_{\nu b})^a  \nonumber \\
        &+ \gamma^{1/2} [D_{\mu}^{a b} ( \phi - \bar{\phi})^b_{\mu a} -
        (D_{\mu} c \times \bar{\omega}_{\mu a})^a ] - \gamma D (N^2-1);
\end{eqnarray}
where $\gamma$ is
the so-called `Gribov parameter'. $(A \times B)^a$ is defined to be
$gf^{abc}A^b B^c$. The horizon condition can be reexpressed as
\begin{equation}
        \label{local_condition}
        \left< \left(g A_{\mu} \times ( \phi_{\mu a} - \bar{\phi}_{\mu a}) \right)^a-(gD_{\mu} c \times \bar{\omega}_{\mu a})^a \right> = 2 \gamma^{1/2} (N^2 - 1) D . 
\end{equation}
The horizon condition can be obtained via an exact, implicit
formula for the eigenvalues of the FP operator. This is done by splitting the
FP operator into two contributions:
\begin{equation}
        \mathcal{M}^{ac} \equiv \mathcal{M}_0^{ac} + \mathcal{M}_1^{ac} = -\partial^2
        \delta^{ac} - g f^{abc} A^{b}_{\mu} \partial_{\mu}  \; . \nonumber 
\end{equation}
Notice we've used the transversality of the gauge field to commute the
derivative through $A_{\mu}$. One can then solve the eigenvalue problem
at low $k$ to see how the degenerate space of eigenmodes of the Laplacian
splits and eventually crosses through $\lambda = 0$. This approach will be the
one taken here to find the finite-temperature horizon condition.

An altogether different way of implementing this constraint was later
found by Maggiore and Schaden (MS) \cite{Maggiore:1993wq}. They simply added a BRST
quartet of fields,
\begin{align}
        s \phi &= \omega \; \; \; \; \; \; \; \; s \omega = 0 \\
        s \bar{\omega} &= \bar{\phi} \; \; \; \; \; \; \; \; \; s \bar{\phi} = 0
        \nonumber .
\end{align}
which is shown in \cite{Vandersickel:2012tz} to not effect physical states.
They added to the action an s-exact term, $s(\partial \bar{\omega} D \phi)$,
made up of these fields. Next they hypothesized a peculiar,
non-translationally invariant field redefinition, parametrized by a
dimensionful quantity, $\gamma$, the same as the Gribov parameter above and
related to the Gribov mass, 
\begin{align}
        \phi^a_{\mu b}(x) &\rightarrow \phi^a_{\mu b}(x) - \gamma^{1/2} x_{\mu} \delta^a_b
        \nonumber \\    
        \bar{\phi}^a_{\mu b}(x) &\rightarrow \bar{\phi}^a_{\mu b}(x) +
        \gamma^{1/2} x_{\mu} \delta^a_b \nonumber \\
        b^a (x) &\rightarrow b^a(x) + i \gamma^{1/2} x_{\mu} f^a[\bar\phi(x)] \nonumber \\
        \bar{c}^a (x) &\rightarrow \bar{c}^a(x) + i \gamma^{1/2} x_{\mu} f^a[\bar\omega(x)] \nonumber,
\end{align}
where $f^a[X] \equiv f^{abc} \; X^b_{\mu c}(x)$.
The value of this parameter was shown to be non-zero at the minimum of the
effective action, thus they obtained a lower vacuum around which to define
their perturbative fields. Demanding that $\gamma$ be defined by minimizing the
effective action is equivalent to the horizon condition (\ref{condition}) and, remarkably,
this procedure yields \emph{exactly} the GZ action.\footnote{We'd like to point
        out that this field redefinition has explicit $x_{\mu}$ dependence and
        hence doesn't vanish at infinity. As in the case of the Higgs VEV,
        this takes us to a different Hilbert space. For a 
        critique of this procedure see \cite{Dudal:2008sp}.} After integrating
out the new ghosts, one arrives back to the non-local action,
which is the integral representation of limiting configuration space to
the Gribov region ($\mathcal{D} A \rightarrow \mathcal{D} A \;
\Theta(H(A))$.) Not only does this remarkable self-consistency
bolster our confidence in the procedure, but it also sheds light on the
explicit (but soft) BRST breaking in the GZ action. From the point of
view of the MS shift, BRST is spontaneously broken only in the ghost
sector \cite{Schaden:2014bea}, which allows one to still use cohomology
to construct the physical Hilbert space. This is an essential
ingredient to quantizing a gauge theory and we would like any
implementation of the Gribov horizon at finite temperature to have this
feature. 

One can see however, that a naive `non-compact time to compact time' procedure
will not be consistent with the MS field redefinition, which depends linearly
on time for the temporal components of the ghosts. The fields that were once
periodic in $\tau \rightarrow \tau + \beta$ will no longer be periodic post MS
shift, but increase linearly with $\tau$ and become multivalued in a manner
akin to the natural logarithm function in the complex plane. 

In the next section we'll see how properly implementing exact degenerate theory techniques with a compact direction leads to a well defined MS shift.

\begin{figure}
        \begin{center}
                \includegraphics[height=4cm]{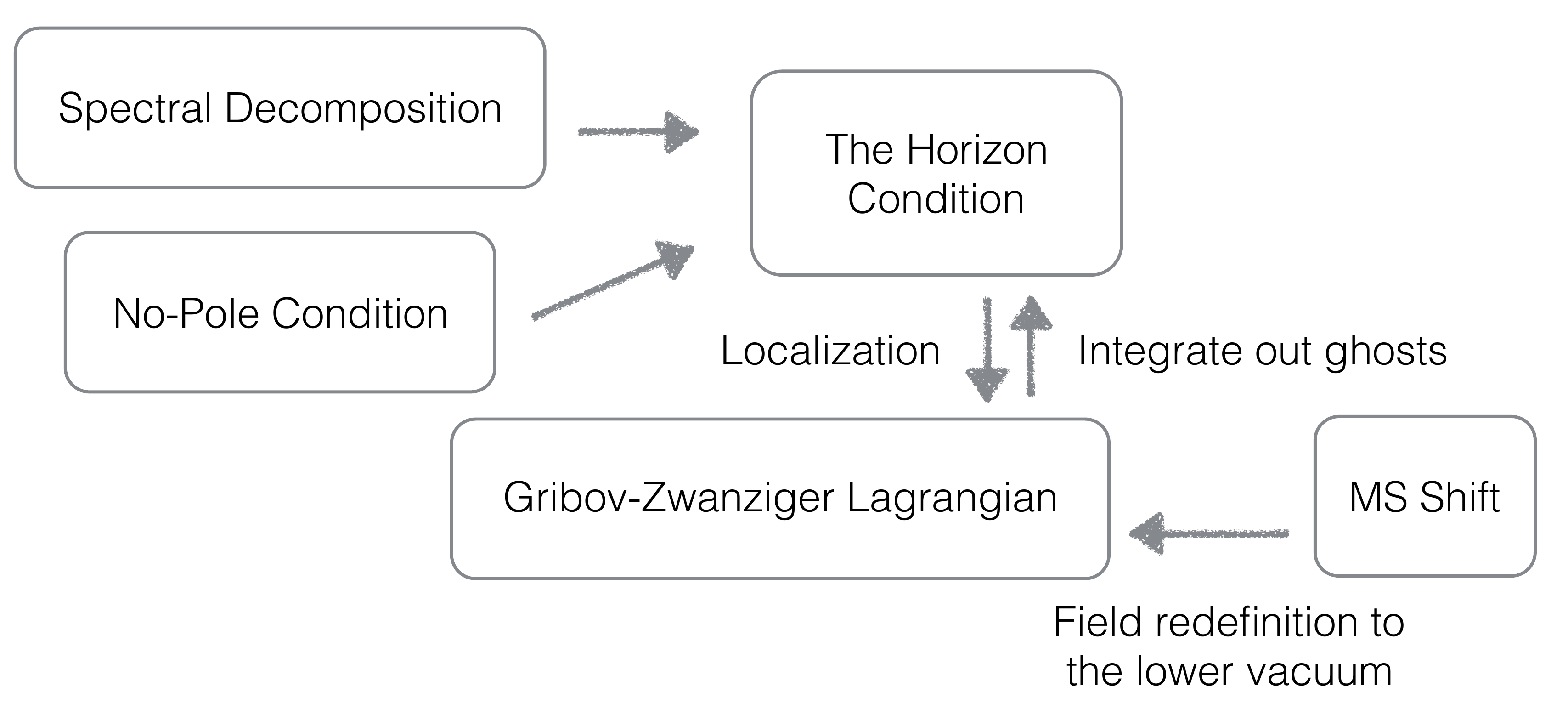}
                \caption{A graphical map of how to define the
                Gribov region.}
                \label{fig:history}
        \end{center}
\end{figure}

\section{The Horizon at Finite Temperature}%

First, let us recall a bit of the discussion in the Introduction. Remember that
at high temperatures, the theory is effectively $D-1 = d$ dimensional due to
the decoupling of the non-zero Matsubara modes whose excitation energy goes like $2
\pi n T$. The modes of the gauge connection which can feel this compact
direction have a finite wavelength and thus for them, $\mathcal{M}_0$
dominates over the perturbation $\mathcal{M}_1$ by a finite, positive value, and the restriction to the
Gribov region has no effect. Since the horizon, $\partial \Omega$, should most
strongly impose constraints on the IR degrees of freedom (where the Laplacian
is small), we should expect that the Gribov procedure should most strongly affect the
IR fields, namely $A_i$, not $A_0$.\footnote{Here we speak only of the non-instantaneous part of $A_0$.  The instantaneous part is long-range, and strongly enhanced in the infrared.} Let's see how this works.

We first start in a finite volume, $V=L^d$, and treat $\mathcal{M}_1$ as a
perturbation. There the spectrum can be well approximated by the spectrum of
Laplacian, $\lambda_k = k^2$, where $k$ is the wave number of the unperturbed
eigenmodes. The lowest value of this eigenvalue at finite volume is $(2 \pi / L)^2$. We then turn on the perturbation adiabatically and
simultaneously take the spatial infinite-volume limit. Due to the fact that the color
structure of the perturbation is completely antisymmetric and thus tracesless,
there are at least some negative eigenvalues of $\mathcal{M}_1$. Because of
this, as we ramp up the perturbation, the spectrum will lower in some color
directions, on top of the fact that the whole spectrum is lowering uniformly as
we take the infinite-volume limit. What is crucial here, is that
the higher Matsubara frequencies will never be the first modes to cross since
they are strictly greater than the zero mode by an amount that remains finite
in the infinite-volume limit, being dependent on the size of the compact
direction, which is only a function of temperature\footnote{see Appendix
        \ref{app:crossing} for a brief discussion about level crossing.}. Thus
when looking at the implicit formula for the eigenvalues, which
schematically takes the form of a scattering problem
\begin{equation}
        \lambda = \lambda_0 + \langle \mathcal{M}_1 \rangle + \sum_{states}
        \langle k | \mathcal{M}_1 \frac{1}{\mathcal{M} - \lambda} \mathcal{M}_1
        | k \rangle , \nonumber 
\end{equation}
we need only consider states with $k_{\mu} = (0,k_i)$ because these modes will
determine the location of the horizon.

In an accompanying paper \cite{Cooper:2015}, we work out the details of this
procedure in the general case. To summarize, the first order perturbation
vanishes since $\mathcal{M}_1$ is traceless. The location of the horizon
($\lambda$ = 0) reduces the implicit eigenvalue equation for what happens to an
eigenvalue that came out of the degenerate subspace of the Laplacian, labeled
by $|\vec{k}|$, down to the form
\begin{equation}
        \lambda_{|k|}(gA) = 0 = k^2 \left( 1 - \eta^{-1} \sum_{\vec{k};|\vec{k}|}
\hat{k}_i \hat{k}_j K^{aa}_{ij}(\vec{k};gA) \right)  \; , \nonumber
\end{equation}
where $\eta = (N^2 - 1) \Sigma_{|\vec{k}|;\vec{k}} 1$ is the degree of
degeneracy of the eigenvalue that we're perturbing due to the cubic symmetry of
the Laplacian. The summation notation means sum over all $\vec{k}$'s with norm
given by $|\vec{k}|$. The expectation value of the regularized horizon
function, $K^{ab}_{ij}(\vec k, g A)$ given by 
\begin{equation}
        K^{ab}_{ij}(\vec{k};gA) \equiv V^{-1} \int d^Dx \; d^Dy \; \mathrm{exp}[i\vec{k} \cdot
        (\vec{y} - \vec{x})] D_i^{ac}(x) D_j^{be}(y)(\mathcal{M}^{-1})^{ce}(x,y)
        \; . \nonumber 
\end{equation}
is the correlator introduced by Kugo and Ojima, $\left< D_i \bar{c}^a D_j c^b
\right>$. 

\begin{figure}
        \begin{center}
                \includegraphics[width=6cm]{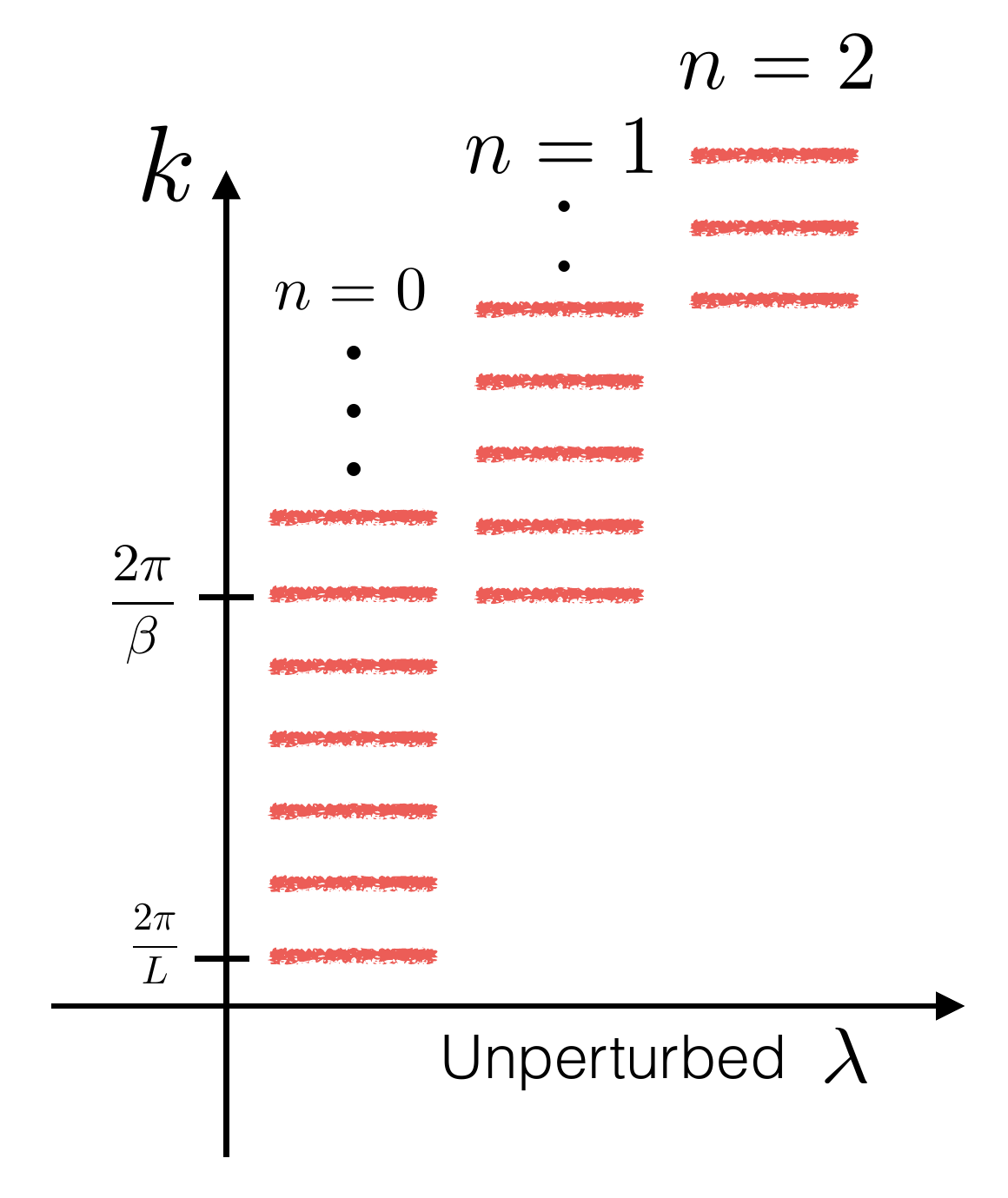}
                \caption{A sketch of the spectrum of $\mathcal{M}_0$ at finite
                temperature}
                \label{fig:levels}
        \end{center}
\end{figure}
The small $k$ limit of this expression, which is of interest to find the
\emph{first} eigenvalue to cross $\partial \Omega$, is discussed in gory detail
in \cite{Cooper:2015}. What is important to us is the fact that the indices of
the regularized horizon function are \emph{not} over $\mu$ and $\nu$ but rather
$i$ and $j$.  In the accompanying paper this was because the discussion was
about Coulomb gauge, but for our finite-temperature discussion, this is because,
when the operator $\mathcal{M}_1$ acts on the states of interest, it brings out a
factor of $k_i$, not $k_{\mu}$ because the zero Matsubara mode will be the first
to cross the horizon. Thus even when considering Landau gauge, the horizon
condition still takes this form. A change of gauge changes only the form of
$\mathcal{M}$, and for the Landau gauge is still given by $\mathcal{M} = - D_\mu \partial_\mu$.

This gives the following horizon condition
\begin{equation}
        \left\langle \int d^Dx d^Dy \; D_i^{ab}(x) D_i^{ac} (\mathcal{M}^{-1})^{bc}(x,y)
        \right\rangle = V d (N^2 - 1).
\end{equation}

\subsection{The Local Action at Finite Temperature}%

Above, we saw how using the horizon condition at finite
temperature required one to not only compactify time as usual, but also to
alter the horizon condition from $D_{\mu} D_{\mu}$ to $D_i D_i$. How does this
new change manifest itself in the local, GZ form of the action? 

When introducing auxillary ghosts to localize this condition in Landau gauge,
we typically give them a Lorentz structure similar to the covariant derivative
structure in the horizon condition because in the end we'd like to make Lorentz
scalars of these objects (see the term $D_{\mu} (\bar{\phi}_{\mu} - \phi_{\mu})$ in $\mathcal{L}_{\rm GZ}$, where the color structure is contracted treating $D$
and $\phi$ in the adjoint representation.) Our results imply that in the GZ
action one should use auxillary ghosts which transform according to $SO(d)$
instead of $SO(D)$ (for the Euclidean theory of course, $SO(1,d)$ in Minkowski
space.) yielding
\begin{eqnarray}
\mathcal L_{GZ} &= \frac{1}{4} F^2_{\mu \nu} + i \partial_{\mu} b \cdot
        A_{\mu} - i \partial_{\mu} \bar{c}^a ( D_{\mu} c)^a + \partial_{\mu}
        \bar{\phi}^a_{ib} (D_{\mu} \phi_{i b})^a - \partial_{\mu}
        \bar{\omega}^a_{i b} \cdot (D_{\mu} \omega_{i b} + D_{\mu} c \times
        \phi_{i b})^a  \nonumber \\
        &+ \gamma^{1/2} [D_i^{a b} ( \phi -
        \bar{\phi})^b_{i a} - (D_i c \times \bar{\omega}_{ia})^a ] - \gamma
        d (N^2-1);
\end{eqnarray}
This means even in Landau gauge, the local horizon
condition (\ref{local_condition}) becomes
\begin{equation}
        \left< g f^{a b c} A_{i}^a \left( \phi^{b c}_{i} - \bar{\phi}^{b c}_{i}
        \right) - (D_i c \times \bar{\omega}_{i a})^a \right> = 2 \gamma^{1/2} (N^2
        - 1) d . \nonumber
\end{equation}
According to the standard lore, at high temperatures $A_0$ is parametrically
more more massive than $A_i$ and decouples. The ratio of masses between $A_0$
($m_{\rm Debye} \sim gT$) and $A_i$ ($m_{\rm magnetic} \sim g^2 T$) vanishes like the
coupling as a function of temperature. Recall in Coulomb gauge, the
instantaneous part of the $\left<A_0A_0\right>$ propagator is bounded from
below by the Wilson potential that is shown via lattice to be linearly rising.
Thus in Coulomb gauge this decoupling appears in the non-instantaneous
contribution, responsible for Debye screening.  This means we expect the
difference between our result and the naive construction of the GZ action in
Landau gauge at very high temperature to be a small perturbation. At
intermediate temperatures however, this can have a sizable effect on performing
perturbative calculations with the GZ action. 

The most remarkable feature of this simple result is that it is consistent with
the well defined MS shift at finite temperature, not affecting the periodicity
conditions imposed by computing the partition function. Specifically
\begin{align}
        \phi^a_{i b}(x) &\rightarrow \phi^a_{i b}(x) - \gamma^{1/2} x_{i} \delta^a_b
        \nonumber \\    
        \bar{\phi}^a_{i b}(x) &\rightarrow \bar{\phi}^a_{i b}(x) +
        \gamma^{1/2} x_{i} \delta^a_b \nonumber \\
        b^a (x) &\rightarrow b^a + i \gamma^{1/2} x_i \; f^{ab}_{\; \; \; c}
        \; \bar{\phi}^c_{i b} (x) \nonumber \\
        \bar{c}^a (x) &\rightarrow \bar{c}^a + i \gamma^{1/2} x_i
        \; f^{ab}_{\; \; \; c} \; \bar{\omega}^c_{i b}(x) \nonumber.
\end{align}
We take the perspective that the action with spontaneously broken, as opposed
to soft BRST breaking, is primary because in that case we believe we know how to define physical states and observables \cite{Schaden:2014bea}.

\section{Applications for Finite Temperature}%

A number of authors have used the GZ action to extend our knowledge about finite-temperature Yang-Mills further into the non-perturbative
regime \cite{Cucchieri:2007ta, Maas:2005ym, Canfora:2013kma}. Our perspective in
light of the altered GZ action is that it is convenient to use Coulomb gauge. The
Coulomb gauge is a unitary gauge which has the same symmetries as the thermal
bath, namely spatial isotropy, but not invariance with respect to boosts. Also,
as previously mentioned, the horizon condition should be purely an IR effect,
and in Coulomb gauge, the full horizon condition only involves the spatial fields which
don't decouple in the high temperature limit. If one is worried about perturbative
renormalizablity, much progress has been made to show that infinities cancel
\cite{Andrasi:2015pha}, however a rigorous proof of existence of perturbation
theory in Coulomb gauge is still lacking. In the following, we'll discuss what
to expect from using the GZ action at finite temperature regarding the magnetic
mass and large N.

\subsection{Coupling Constant Expansion, the Magnetic Mass and Condensates}%

As was previously mentioned, the free energy of a thermal bath of gluons is not
perturbatively calculable in a coupling constant expansion the same way it is
for QED. With Yang-Mills coupling $g$, one starts to run into infrared
divergences when attempting to calculate $g^6$ contributions. To give a
physical picture for the origin of this non-perturbative physics, Gross et al
in their '81 review on finite-temperature QCD \cite{Gross:1980br} showed
that if one were to attempt to estimate the contribution to the free energy of
topologically unstable Wu-Yang monopoles, one would find an $O(g^6)$
contribution to the free energy. Their back-of-the-envelope calculation was
intended to bolster the intuition that non-perturbative effects from the
effectively $(D-1)$ dimensional gauge theory at high temperature become
unavoidable at a high enough order in $g$. 

In the presence of new Bose fields in the GZ action, we should make sure the
analysis about monopoles is still applicable. That is, we need to show that the
equations of motion for the gauge field are not affected by the auxillary
ghosts. This can be easily demonstrated for the ghosts introduced by
the Fadeev-Popov procedure. Since the equations of motion are gauge dependent,
let us consider the following gauge-independent observable
\begin{equation}
        G \equiv \left\langle \frac{\delta S}{\delta A_{\mu}^a(x)} \frac{\delta
        S}{\delta A_{\mu}^a(x)} \right\rangle . \nonumber
\end{equation}
The contribution to the Lagrangian from gauge fixing to the Landau gauge is
simply
\begin{equation}
        \mathcal{L}_{\mathrm{FP}} = \mathrm{s}\left( \partial'_{\mu}
                \bar{c} A_{\mu}
        \right) , \nonumber
\end{equation}
where the prime derivatives stand for $\partial'_{\mu} = (\alpha \partial_0,
\partial_i)$ which allows us to cover all interpolating gauges from Landau
($\alpha = 1$) to Coulomb ($\alpha = 0$) and $s$ is the usual BRST
transformation. After performing the variation, one obtains
\begin{equation}
\label{FP_part}
        \frac{\delta S}{\delta A_{\mu}^a} = D^{ab}_{\lambda} F^b_{\lambda \mu}
        + \partial'_{\mu} b^a - g \partial'_{\mu} \bar{c}^b f^{b a
        c} c^c . 
\end{equation}
G can then be written
\begin{eqnarray}
        G &=& \left\langle (D_{\lambda} F_{\lambda \mu})^2 \right\rangle +
        \left\langle s\left\{ \partial'_{\mu} \bar{c} \cdot g
                (\partial'_{\mu} \bar{c} \times c) + D_{\lambda} F_{\lambda
        \mu} \cdot \partial'_{\mu} \bar{c} \right\} \right\rangle \nonumber \\
        &=&\left\langle (D_{\lambda} F_{\lambda \mu})^2 \right\rangle , \nonumber
\end{eqnarray}
with the last line following from the fact that BRST is unbroken by the
Fadeev-Popov contribution to the action. The following questions remain: can
the equivalent observable, $G$, in GZ theory be written as the Yang-Mills
squared equation of motion plus an s-exact term, and secondly, is the
expectation value of this s-exact contribution zero? 

The pre-MS shifted contribution to the Lagrangian takes the following simple
form (suppressing color indices and Lorentz indices on auxiliary ghosts for
readability)
\begin{equation}
        \mathcal{L_{\rm GZ}} = s(\partial'_{\mu} \bar{\omega} D_{\mu} \phi ).
        \nonumber 
\end{equation}
Thus the contribution to the equations of motion are
\begin{eqnarray}
        \frac{\delta S_{\rm GZ}}{\delta A_{\mu}} &=& -g \partial'_{\mu} \bar{\phi}
        \times \phi + g \partial'_{\mu} \bar{\omega} \times \omega - g^2
        (\partial'_{\mu} \bar{\omega} \times \phi) \times c \nonumber \\
        &=& s(-g \partial'_{\mu} \bar{\omega} \times \phi) + g
        (-g \partial'_{\mu} \bar{\omega} \times \phi) \times c . \nonumber
\end{eqnarray}
The second equality is written to reflect the equivalent form of equation
\eqref{FP_part} above, which can be rewritten
\begin{equation}
        \frac{\delta S_{\mathrm{FP}}}{\delta A_{\mu}} =
        s(\partial'_{\mu} \bar{c} ) + g
        (\partial'_{\mu} \bar{c}) \times c . \nonumber
\end{equation}
Now the equations of motion take the form
\begin{equation}
        \frac{\delta S}{\delta A_{\mu}} = D_{\lambda} F_{\lambda \mu} + s
        X_{\mu} + g X_{\mu} \times c , \nonumber
\end{equation}
with
\begin{equation}
        X_{\mu} \equiv \partial'_{\mu} \bar{c} - g \partial'_{\mu} \bar{\omega}
        \times \phi . \nonumber
\end{equation}
It can then be easily checked that 
\begin{equation}
        G = \left\langle (D_{\lambda} F_{\lambda \mu})^2 + s\left\{ 2 D_{\lambda} F_{\lambda
        \mu} X_{\mu} + X_{\mu} (s X_{\mu} + g X_{\mu} \times c)
        \right\} \right\rangle , \nonumber
\end{equation}
with $X_{\mu}^a$ being an arbitrary color and Lorentz vector. Thus, as
expected, soliton solutions to Yang-Mills will indeed be solutions to the
classical equations of motion to GZ theory, provided the s-exact contribution
to the equations of motion has a vanishing vacuum expectation value. This
condition is trivial for the FP action because the vacuum is BRST invariant.
With the GZ action, the vacuum breaks BRST spontaneously and the vanishing of
the expectation value of an s-exact operator is only to be expected for the
restricted class of observables mentioned in \cite{Schaden:2014bea}, \emph{and}
when the horizon condition is imposed. Checking this by hand was done for the
stress-energy tensor of the theory, and the term relevant in constructing the
Kugo-Ojima confinement criterion argument in \cite{Schaden:2014bea}, but for the
equations of motion, we will have to simply make use of the unproven conjecture
that observables do not break BRST symmetry. Checking the vanishing
of this term explicitly will be left for future work and would provide a very
non-trivial check of the expectation that BRST holds exactly when needed.

The estimation of the effect of 3-dimensional monopole configurations can be
interpreted as a magnetic mass for the spatial gluons. Karabali and Nair give a
thorough treatment of Yang Mills theory in 3 spacetime dimensions
\cite{Karabali:1998yq}. Similarly to how holomorphic coordinates aid one in
calculations of electrodynamics in 2+1 dimensions, holomorphic coordinates have
also proven to be useful in the Karabali-Nair formalism. As well as treating
space as a single complex variable, they also explicitly calculate the
functional measure for a change of field space coordinates from the gauge
fields to gauge-invariant quantities.  In this way they circumvent the Gribov
ambiguity because the description no longer has a gauge redundancy.  This
allows them to estimate certain non-perturbative objects, like the magnetic mass in
terms of the dimensionful 3D coupling. In \cite{Nair:1998es} Nair estimates for
the magnetic mass, $m = g^2 N / 2 \pi$ where $g$ is now the three dimensional
coupling constant which is related to the 4-D one by $g_4^2 T = g_3^2$ in the
high temperature dimensionally reduced picture. 

Jackiw et al in \cite{Jackiw:1995nf} began a program of calculating the
magnetic mass by simply adding and subtracting a mass term to the Lagrangian.
This trivial operation is then interpreted as altering the free propagator to
be a massive one (curing IR divergences) and adding a quadratic interaction. By
demanding that the value of this mass be self-consistent, one seeks to solve
the ``Gap Equation", which is the statement that the value for the added mass
should lead to 0 corrections of the gluon self-energy. Others continued this
program to 2 loop (\cite{Cornwall:2015lna},\cite{Eberlein:1998yk}) and obtained
values somewhat off from Nair's estimate, namely $g^2 N / 6 \pi$. In the
formulation of this technique, nothing guarantees that solving the gap equation
at higher loop order leads to a parametrically better approximation to the
magnetic mass. A cross check with another technique is needed.

The new and improved propagators of the GZ action come equipped with an
infrared cutoff that naively would cure the divergences that plague
finite-temperature QCD. Furthermore, it has been shown that the horizon
condition yields a Gribov mass on the order of $g^2 T$, precisely that of the
magnetic mass \cite{Zwanziger:2006sc}. The relation between the Gribov mass and
the magnetic mass supports the hypothesis that the non-trivial topology of the
space of gauge orbits is a crucial aspect to understanding the phenomenology of
non-abelian gauge theories. 

This doesn't solve the problem however; the Gribov mass would still receive
contributions from higher order gluon self-energy diagrams, and be just as
difficult to calculate as the magnetic mass or $\Lambda_{QCD}$. One could argue
that all we've done is moved the non-perturbative physics from the fully
resummed magnetic mass to the value of the Gribov mass. This could be
advantageous however because it gives us another way to parameterize our
ignorance. The value of the Gribov mass is given by the horizon condition, a
non-perturbative formula which might be easier to estimate than the
perturbative expansion for the magnetic mass. 

For example, the GZ approach yields an alternative perturbation scheme as a
cross check to the results obtained by the program mentioned above. Instead of
calculating the mass gap in the three dimensional theory, we perturbatively
solve for the Gribov mass in the full 4-dimensional theory. This scheme is a
dual perturbation, where one calculates the contributions to the free energy to
lowest order in $g$ treating the Gribov mass, $m$ as $O(g^0)$. Then you solve
the horizon condition to that order and find the $g$ dependence of the Gribov
mass. You then plug in that ansatz for $m$ and move to the next order in the
coupling constant expansion, and proceed from there. In \cite{Zwanziger:2006sc}
this procedure was used to find the first-order high-temperature contribution
to the non-perturbative scale of the QGP equation of state, and in Appendix
\ref{app:commute} we show that this procedure is consistent with doing the same
calculation in the dimensionally reduced theory as a non-trivial cross check.
The result to lowest order is $m/T = g^2 N / 6 \sqrt{2} \pi$, not far from the
estimate of Cornwall et al \cite{Cornwall:2015lna} for the magnetic mass given
above. 

The Gribov mass is also relevant for physics regarding condensates.  For
example, as is nicely reviewed in \cite{Mathieu:2008me}, the dynamical mass of
the dressed gluon propagator can be used to calculate the glueball spectrum,
which can be related to the gluon condensate, $\langle F_{\mu \nu}^a F^{\mu
\nu}_a \rangle$. The gluon mass spectrum has been fit quite well in the context
of refined GZ theory, in \cite{Sorella:2011tu} which uses the GZ action with
phenomenological corrections. An old result \cite{Collins:1976yq} relates the
quantum anomaly, $\mathcal{A}$, in the trace of the stress energy tensor of QCD
to the beta function and the value of the gluon condensate,
\begin{equation}
        \mathcal{A} = \frac{\beta(g)}{2g} \langle : (F^a_{\mu \nu})^2 : \rangle \; . \nonumber
\end{equation}
Similarly, one can calculate the trace anomaly in the context of GZ theory,
which will be depend on the Gribov mass. The tree level result from
\cite{Schaden:2014bea} yields
\begin{equation} 
        \mathcal{A} \sim -(N^2-1) \frac{3 m^4}{4 (2 \pi)^2} \; . \nonumber
\end{equation} 
Again, this gives a dual perturbation for calculating the gluon condensate,
simultaneously expanding the trace anomaly and the horizon condition as above.

\subsection{Large-N}%

Finally, we would like to mention that a potential research direction
would be to use the GZ action at finite temperature to provide a
different approach to large-N QCD. Thinking about the effects of a gauge-fixing
ambiguity resulting from non-trivial topologies has been studied in the past in
a much better understood setting. Haber et al \cite{Haber:1980uy} studied the
$\mathbb{C}P^{N-1}$ model with unconstrained fields and noticed that a gauge
redundancy existed due to precisely the pole of $\mathbb{C}P^{N-1}$ which gives
this sigma-model its non-trivial topology. Only a subset of fields in this
model have this residual gauge freedom\footnote{Those authors referred to this as a sort of Gribov ambiguity as well \cite{Haber:1980uy}}, but in the
large-N limit, this set of measure-zero gets emphasized, meaning the expectation-value of the fields is precisely the value at which the ambiguity resides. This
is similar to how in the infinite-volume limit, the horizon condition puts the
measure of configuration space on the horizon which, as shown in
\cite{Cooper:2015}, is also a geometrically priveleged set of configurations.
Haber et al found that this special set of fields led to a singularity in the
Hamiltonian, not unlike the one found in \cite{Cooper:2015}, and that this
singularity is responsible for the IR confinement of the theory. 

The Gribov mass could be more than another way to parameterize our ignorance
about $(D-1)$ dimensional confining theories, it could perhaps lead to a better $1/N$
expansion. From Nair's reformulation of 3-D QCD, the Gap equation, and the
Horizon condition approach, it seems perturbatively stable that the magnetic
mass is exactly proportional to N. Gracey's two-loop results for the Horizon
condition \cite{Gracey:2005cx} also suggest that this result persists at higher order.

As we show in the Appendix \ref{app:commute}, the Gribov
mass gives another dimensionful parameter in the dimensionally reduced theory.
Thus in the case of a $D=4$ gauge theory at high temperature, we have two
dimensionful quantities: the 3-dimensional coupling constant given by
$g_3^2 = g_4^2T$ and the Gribov mass, $m$. The ratio between these two is of
order $1/N$. Thus we now have a small parameter in the theory to
potentially reorganize the perturbation theory. Recall in the large-N limit
$g^2$ goes to zero, which corresponds to large T which is exactly where the 3
dimensional description becomes valid. An interesting remark, is that
this doesn't mean that the effect of the Gribov horizon decouples in the large-N
limit. The parameter in the MS shift is related to the Gribov mass by
\begin{equation}
\label{gammatom}
    \gamma^{1/2} = \frac{m^2}{\sqrt{N} g} \; . 
\end{equation}
In terms of the 't Hooft parameter, $x = g^2 N$ which remains fixed in the
large-N limit, $m \sim x$ (see Appendix \ref{app:commute}) and $\gamma \sim x^3$.

\section{Conclusion}%

In this chapter we've revisited the analysis of finding the Gribov region using
degenerate perturbation theory; we argue that only the zero Matsubara mode
must be considered. This intuitive result dovetails nicely with the
assumption that the Gribov procedure is a correction due to the non-trivial
topology of the equivalence classes of gauge orbits, and only IR physics should
feel this non-triviality. This correction to the horizon condition has
implications for what form of the GZ action one should use in various gauges.
The correct form of the GZ action happens to coincide with the only well-defined MS shift procedure at finite temperature which is a nice self-consistency check for the theory, since the MS shift allows us to define
physical states despite the apparent BRST breaking of the usual GZ action.
Since the Gribov mass provides an infrared cutoff, it could play an important
role in calculating the value of the magnetic mass. The extent of the role of
the Gribov ambiguity in illuminating non-perturbative physics is still unknown at
this point, but certainly it gives us an alternative approach to understanding
these effects. 

\smallskip
{\bf Acknowledgements}\\
We recall with pleasure stimulating conversations about this work with Geoff
Ryan, Luke Underwood, Martin Schaden and Eliezer Rabinovici. 
\appendix
\section{Commuting of the High Temperature Limit} \label{app:commute}%

Above we described a procedure for calculating the Gribov
mass perturbatively in the high-temperature limit. This was done by truncating
the horizon condition at a certain order of the coupling, $g$, while treating
the Gribov mass, $m$, as $\mathcal{O}(g^0)$. After summing over Matsubara
frequencies, we can then look at the horizon condition in the limit
$T\rightarrow \infty$ to get a closed for expression for $m(g,T)$. However,
there is a completely different way of taking the high temperature limit. At
high $T$ (low $\beta$), the compact direction becomes vanishingly thin, and
what results is an effectively $d$ dimensional, zero temperature euclidean
field theory with $g_d^2 = T g_{d+1}^2$. In this appendix, we shall calculate
the horizon condition to low order in arbitrary euclidean dimensions, and show
that the 3D result, with $g_3^2 = g_4^2 T$ yields the exact expression derived
for the finite temperature 4-dimensional Gribov mass. Thus the perturbative
procedure described above for imposing the horizon condition commutes with the
order in which you take the high temperature limit and one can avoid matsubara
sums altogether at high temperature by using the dimensionally reduced theory.

For arbitrary euclidean space dimension, d, the functional integral up to
quadratic terms is given by
\begin{equation}
        \mathcal{Z} = e^W = \int dA \, db \, dc \, d\bar{c} \, d\phi \,
        d\bar{\phi} \, d\omega \, d\bar{\omega} \; \; e^{-S} \nonumber,
\end{equation}
with 
\begin{align}
        S = \int d^dx & \left( \frac{1}{4} F_{0 \mu \nu}^a F_{0 \mu \nu}^a + i
        \partial_{\mu} b A_{\mu} - i \partial_{\mu} \bar{c}
            \partial_{\mu} c + \partial_{\mu} \bar{\phi}_{\nu} \cdot
            \partial_{\mu} \phi_{\nu} + \partial_{\mu} \bar{\omega}_{\nu} \cdot
            \partial_{\mu} \omega_{\nu} + \nonumber \right. \nonumber 
            \\ 
            &g  \left. \gamma^{1/2} f^{abc} 
            A_{\mu}^b (\phi_{\mu} - \bar{\phi}_{\mu})^{ca} - \gamma d (N^2-1)
            \frac{}{}\right) \nonumber
\end{align}
where $\gamma$ is given by (\ref{gammatom}) and $F_0$ means $F$ restricted to
the linear terms in $A$. Note, here $d$ is arbitrary and $\mu, \nu$ range from
$1,\ldots,d$. We will use it for $d=3$ in the end, so in that case $\mu, \nu$
would be $i,j$ to be consistent with what's presented above. Integration over
the fermi ghosts $c$, $\bar{c}$ and the Nakanishi-Lautrup field b implements
the Landau gauge condition, leaving only the transverse part of $A_{\mu}$. One
can also perform the field redefinition \cite{Maggiore:1993wq},
\begin{align}
        &\phi^{ab}_{\mu} \rightarrow \phi^{ab}_{\mu} + g \gamma^{1/2}
        (-\partial^2)^{-1} f^{cab} A^c_{\mu} \nonumber \\
        &\bar{\phi}^{ab}_{\mu} \rightarrow \bar{\phi}^{ab}_{\mu} - g \gamma^{1/2}
        (-\partial^2)^{-1} f^{cab} A^c_{\mu} \nonumber . \nonumber
\end{align}
This will diagonalize $\phi$'s kinetic term at the cost of adding to the action
the term
\begin{equation}
        g^2 \gamma f^{abc} A^a_{\mu} (-\partial^2)^{-1} f^{dbc} A^d_{\mu}
        . \nonumber
\end{equation}
The inverse determinant which results in integrating out the shifted bose ghosts
cancels the determinant from integrating out the auxillary fermi ghosts and
we're left with
\begin{align}
        S =  \int d^dx \left( \frac{1}{2} \partial_{\mu} A^{tr}_{\nu} \cdot \partial_{\mu}
        A^{tr}_{\nu} + m^4 A^{tr}_{\mu} \cdot (-\partial^2)^{-1}
        A^{tr}_{\mu} - \gamma d (N^2 - 1) \right) . \nonumber 
\end{align}
Focusing on the quadratic part of the action, $S_0$, we have
\begin{align}
        e^{W_0} =  \int dA^{tr} \, e^{-S_0} =
        \left( \mathrm{det}^{-1/2}\left[-\partial^2 +
        \frac{m^4}{-\partial^2}\right] \right)^{(N^2-1)(d-1)} \nonumber \\ 
        = \mathrm{exp} \left( -\frac{1}{2}(N^2-1)(d-1)V \int \frac{d^dk}{(2
        \pi)^d} \mathrm{ln}\left[ k^2 + \frac{m^4}{k^2} \right] \right) . \nonumber  
\end{align}
Including the constant part of the action, the free energy, $W$, is then given
by
\begin{equation}
        W = W_{-2} + W_0 = \nonumber \\ \frac{(N^2-1) d m^4 V}{2 N g^2} \; - \; \frac{(N^2 -
        1)(d-1) V}{2} \int \frac{d^dk}{(2\pi)^d} \mathrm{ln}\left( k^2 +
        \frac{m^4}{k^2} \right) . \nonumber
\end{equation}
The horizon condition to this order can then be written
\begin{equation}
        0 = \frac{\partial W}{\partial m^4} = \frac{d}{g^2 N} - (d-1) \int
        \frac{d^d k}{(2\pi)^d} \frac{1}{(k^2)^2 + m^4} . \nonumber
\end{equation}
One can then evaluate the integral, yielding
\begin{equation}
        I_d \equiv \int \frac{d^d k}{(2\pi)^d} \frac{1}{(k^2)^2 + m^4} = \frac{1}{m^{4-d}}
        \frac{\pi}{2} \frac{\Gamma(2-d/2)}{\Gamma \left( \frac{1}{2} \left[
        \frac{d}{2} + 1 \right] \right) \Gamma \left( \frac{1}{2} \left[ 3 -
        \frac{d}{2} \right] \right)} . \nonumber 
\end{equation}
$d=3$ gives us
\begin{equation}
        I_3 = \frac{1}{4\sqrt{2} \pi m} . \nonumber
\end{equation}
Thus the horizon condition in 3 dimensions holds for a Gribov parameter given
by
\begin{equation}
\label{gribov_result}
        m = \frac{g_3^2 N}{3 \cdot 2^{3/2} \pi} .
\end{equation}
If we're to believe the high-temperature limit where the 4 dimensional theory
with coupling constant, $1/g_4^2$, is replaced by a three dimensional theory
with coupling, $1/(g_4^2 T) \equiv 1/(g_3^2)$, then we better get the same
dependence of $m$ on $g$ that we found in the high-temperature limit previously
calculated in the strictly 4D theory. Indeed, by taking the three dimensional
result of equation (\ref{gribov_result}) and making the proper replacement, we
have 
\begin{equation}
        m = g_4^2 T \frac{N}{3\cdot2^{3/2} \pi} . \nonumber
\end{equation}
This agrees with equation (94) in \cite{Zwanziger:2006sc}, thus confirming that
taking the high-temperature limit in these two different ways does indeed yield
the same result in this calculational scheme. This effectively allows us to
analyze the high-temperature limit of finite-T GZ theory by doing a standard
calculation in zero temperature euclidean field theory, thereby avoiding the
infinite sum over Matsubara frequencies in the intermediate steps before taking
the limit.

\section{On Level Crossing} \label{app:crossing} %

One important assumption is made when moving from a sum over the set of
eigenvalues of the Fadeev-Popov operator that emerge from the degenerate
subspace of the Laplacian operator when the perturbation,
$gA_{\mu}(x)\partial_{\mu}$, is turned on, to the average of these eigenvalues. That
is, that we can't have a situation where the average is zero, but some of the
individual eigenvalues emerging from that degenerate sub-space could be very
non-zero. This would violate our definition of the Gribov region, as well as
call into question the stability of the ghost kinetic term, which would no
longer be positive definite.

The crux of the argument is that if one takes a generic eigenvalue of the
Laplacian, the splaying of that degenerate subspace into its non-degenerate
spectrum will be sandwiched between the non-degenerate spaces of the higher and
lower eigenvalues of the Laplacian. In the infinite-volume limit, the spectrum
of the Laplacian becomes continuous, therefore the span of that non-degenerate
space becomes infinitely thin, and the average represents any one of the
eigenvalues arbitrarily well. This `sandwiching' caused by the absence of level
crossing is a big assumption, but is well supported by work on early atomic
physics dating back to Wigner and von Neumann \cite{vonNeumann:1929}. 
They show that for a 1 or 2 parameter family of adiabatic perturbations of a
Hermitian operator, levels will generically repel each other unless a symmetry
principle allows otherwise. The symmetries of the unperturbed operator in our
case, the Laplacian, are the generators of $SO(d)$. The FP operator does not
share these symmetries, or for generic $A_i$, any symmetries at all, so its
eigenstates will not fall into different irreducible representations which
allow for generic level crossing. There are two important caveats to this. One,
is the case of accidental crossing. Without an explicit calculation of the full
spectrum this could never be disproved, however the set of gauge fields with
these crossings should have a very small measure and contribute little to the
path integral. The second caveat is more severe, namely that these proofs
about level crossing are typically for finite dimensional matrices, but in
order to justify the concentration of measure on the boundary of the Gribov
region, we must take the infinite volume limit. At this moment we can offer
nothing to ease the skeptics' mind on that issue.

\bibliography{finiteT}

\end{document}